# Direct epitaxial growth of polar (1-x)HfO$_2$-(x)ZrO$_2$ ultra-thin films on Silicon

Pavan Nukala[1,#,*], Jordi Antoja-Lleonart[1,#], Yingfen Wei[1], Lluis Yedra[2,3], Brahim Dkhil[2], Beatriz Noheda[1,*]

1. Zernike Institute of Advanced Materials, University of Groningen, 9747 AG, The Netherlands
2. Laboratoire Structures, Propriétés et Modélisation des Solides, CentraleSupélec, CNRS-UMR8580, Université Paris-Saclay, 91190 Gif-sur-Yvette, France
3. Laboratoire Mécanique des Sols, Structures et Matériaux, CentraleSupélec, CNRS UMR 8579, Université Paris-Saclay, 91190 Gif-sur-Yvette, France

\# equally contributing authors

*Corresponding authors, email: p.nukala@rug.nl, b.noheda@rug.nl

**Abstract**: Ultra-thin Hf$_{1-x}$Zr$_x$O$_2$ films have attracted tremendous interest owing to their Si-compatible ferroelectricity arising from polar polymorphs. While these phases have been grown on Si as polycrystalline films, epitaxial growth was only achieved on non-Si substrates. Here we report direct epitaxy of polar phases on Si using pulsed laser deposition enabled via *in situ* scavenging of the native a-SiOx under ballistic conditions. On Si (111), polar rhombohedral (r)-phase and bulk monoclinic (m-) phase coexist, with the volume of the former increasing with increasing Zr concentration. R-phase is stabilized in the regions with a direct connection between the substrate and the film through the compressive strain provided by an interfacial crystalline c-SiO$_2$ layer., The film relaxes to a bulk m-phase in regions where a-SiO$_x$ regrows. On Si (100), we observe polar orthorhombic o-phase coexisting with m-phase, stabilized by inhomogeneous strains at the intersection of monoclinic domains. This work provides fundamental insight into the conditions that lead to the preferential stabilization of r-, o- and m-phases.

**Introduction:**

Ferroelectric hafnia-based thin-films[1] have by now been established as the most promising materials to realize the potential of ferroelectric phenomena in real devices[2,3]. Their Si compatibility, simple chemistry and unique ferroelectricity, that becomes more robust with miniaturization, is tailor-made for microelectronics, offering ready-made alternatives to conventional ferroelectrics that lack all these attributes[4–12]. Such distinguishing characteristics lead to an upsurge in application-oriented research as well as in curiosity-driven fundamental research to solve questions such as why these materials are capable of sustaining the unconventional ferroelectricity[13–32], how these materials negate the effects of depolarization fields[33,34], and whether such a new type of ferroelectricity can be replicated in other simple oxide systems.

A prominent feature of hafnia-based materials is polymorphism[35]. While the ground state in the bulk $HfO_2$ is a non-polar monoclinic (m-, *P2₁/c*) phase, a plethora of low-volume both polar and non-polar metastable states can be stabilized at ambient conditions via a combination of strategies such as cationic and anionic doping[1,25–27,29,32], thermal and inhomogeneous stresses[36,37], nanostructuring[38], epitaxial strain[16,20,22,23,26,29,39–42], and oxygen vacancy engineering[43,44], all of which can be suitably engineered into thin-film geometries. Based on first-principles calculations[15,39,45-47] at least five polar polymorphs (with space groups *Pca2₁, Cc, Pmn2₁, R3* and *R3m*) can be identified as those that can be experimentally obtained. Owing to its relatively low energy, the orthorhombic (o-) *Pca2₁* phase is widely observed in hafnia-based films grown via atomic layer deposition (ALD)[1,17,24,25], chemical solution deposition (CSD)[28], RF sputtering on Si[18,21] and pulsed-laser deposition (PLD) on selected substrates[19,23,26,31,40–42]. A slightly higher energy rhombohedral (r-) phase (*R3m* or *R3*) has been recently observed on epitaxial $Hf_{1/2}Zr_{1/2}O_2$ films grown on $SrTiO_3$ (STO) [39]. The r-phase is stabilized by a combination of the large surface energy induced internal pressure of the nanoparticles and the substrate-imposed compressive strain. The epitaxial growth of the r-phase enabled the observation of the highest values of spontaneous polarization ($P_s$= 34 µC/cm$^2$) in $HfO_2$-$ZrO_2$ systems, although they showed larger coercive fields than films in the polar o-phase.

So far, polar phases have been successfully obtained via epitaxial synthesis techniques (PLD) on a variety of substrates: perovskites[16,40–42] (including buffered STO on Si (001))[42], fluorites[19,23,26] and hexagonal substrates[48]. However, the advantages gained by epitaxy are offset by the fact that none of these films are grown directly on Si, despite the Si integrability of hafnia-based systems. It is in this context that we explore for the first time the epitaxy of polar polymorphs in (1-x)$HfO_2$-x$ZrO_2$ (HZO (x)) films directly on Si.

Direct epitaxial growth of oxides on Si is complicated by the presence of a very thin amorphous native oxide layer, which prevents the transfer of texture from the substrate to the film. This interfacial layer can be removed prior to deposition through hydrofluoric (acid etching), which lowers the crystalline quality[49,50]. However, the epitaxial growth of yttria-stabilized zirconia (YSZ) on Si (100) is a mature process[51–54], as witnessed by the fact that YSZ-buffered Si (100) substrates are commonly used for the growth of high-$T_c$ superconductors and other functional layers[55–59]. The problem of native oxide in YSZ on Si is solved via an *in situ* scavenging process through Zr (Y), because the formation energy of $ZrO_2$ ($Y_2O_3$) is less than that of $SiO_2$.[50,57,60]. In other words, one (or both) of the following decomposition chemical reactions takes place during the initial stages of YSZ growth enabling a transfer of substrate texture.

$$SiO_2 + Zr \rightarrow Si + ZrO_2 \quad (1)$$

$$2\,SiO_2 + Zr \rightarrow ZrO_2 + 2SiO \uparrow \quad (2)$$

The rest of the YSZ growth follows the template set by the crystalline $ZrO_2$ seed, formed as a result of scavenging the native amorphous oxide, resulting in a very high crystalline quality. Strain relaxation of thicker layers of YSZ happens *via* the regrowth of the a-$SiO_2$ oxide (backward reactions 1 and 2, upon increasing the amount of product), and through other standard epitaxial relaxation processes such as generation of misfit dislocations. Inspired by the success of direct epitaxy of a sister compound YSZ on Si, here we report successful growth of epitaxial polar phases of $HfO_2$-$ZrO_2$ alloys directly on Si (111) and Si (100).

**Experimental Methods:**

PLD was used for the deposition of HZO (x) with x=0.5, 0.7 and 0.85 on *p*-doped Si (111) (resistivity < 0.005 Ω cm), and Si (100) (resistivity < 0.03 Ω cm). Targets of the desired compositions (x=0.5, 0.7, 0.85) were prepared through standard solid-state synthesis starting from powders of $HfO_2$ (99% purity) and $ZrO_2$ (99.5 % purity). A KrF excimer laser (λ=248 nm) was used for target ablation at a fluence of 1.1 J cm$^{-2}$. A base pressure of $10^{-7}$ torr was maintained in the deposition chamber. Target to substrate distance was fixed at 50 mm. HZO layers were deposited at 800°C with the flow of Ar carrier gas (5 sccm) at a process pressure of 0.005 mbar and laser repetition rate of 7 Hz.

The choice of target-substrate distance, as well as the low pressure conditions, ensured a ballistic mode of deposition, preventing the oxidation of atomic species (Hf, Zr ions) in the plasma itself. It is indeed crucial for the interfacial scavenging reactions that species transported to the substrate are Hf, Zr and O ions with no ionized $HfO_2$ and $ZrO_2$ present in the plasma. The 7 Hz repetition rate favors nucleation-dominated kinetics, or in this case, the occurrence of interfacial scavenging at multiple locations, and thus the generation of several (Hf)$ZrO_2$ seeds or growth templates.

On Si(111), HZO (x) films of 10 nm thickness with three different compositions (x=0.5, 0.7, 0.85) were deposited, at a growth rate of 0.7 Å/sec. On Si(100), films of 5, 10 and 20 nm were deposited with composition x=0.7, at a growth rate of 0.9 Å/sec. Thickness was confirmed from both Scanning Transmission Electron Microscopy (STEM) and x-ray reflectivity measurements. Information about global structure, symmetry, phase-mixing and domains was inferred from x-ray diffraction (XRD) with a Cu Kα source. Texture analysis was performed *via* χ-φ (pole figure) scans at 2θ≈30.0° (approximately corresponding to the $d_{\{111\}}$ of the low-volume phases) and at 2θ≈34.5° (approximately corresponding to the $d_{\{200\}}$ of all the polymorphs). The d-spacings of the poles obtained from the χ-φ stereographic projections were more precisely analysed through θ-2θ scans around them, which from here on will be referred to as 'pole-slicing'.

Local structural characterization and phase analysis was performed through STEM imaging at 200 kV (Titan G2 and Themis). STEM images were obtained in both high-angle annular dark field (HAADF) mode, and bright-field (BF) mode. Chemical maps were generated via energy dispersive spectroscopy (EDS) in a four detector ChemiSTEM set-up on the Titan G2 aberration-corrected electron microscope.

**Results and discussion:**

**(1-x)HfO$_2$- xZrO$_2$ on Si (111)**

**Strongly textured (111) films:** Pole figures obtained from films with x=0.5 (Fig. 1a), 0.7 and 0.85 (*Supplementary Fig. S1a*) about 2θ=30.0° look qualitatively similar. In addition to the peak at the center (out-of-plane), there are three poles (P1 to P3) arising from the film at χ~71° separated from each other in φ by ~120°. The weaker poles at χ~71° (Fig. 1a) are from the tail of the substrate peak at 2θ=28.44°. This symmetry is consistent with <111>-oriented films, following the substrate. Quite interestingly, the {111} poles corresponding to the substrate and the film are rotated 180° about the substrate normal.

**Phase coexistence and evidence of r-phase:** On the films with x=0.5, the pole slicing across P1 to P3 clearly shows two peaks at every pole centered with 2θ=28.5° and 31.4° (Fig. 1b), corresponding to $d_{\{111\}}$ and $d_{\{11\text{-}1\}}$ of the bulk m-phase. With the increase of Zr concentration to x=0.7 (arrow in Fig. 1c), a peak corresponding to a low-volume phase starts to appear at 2θ=30.23°, with the majority phase still being monoclinic. A further increase in Zr concentration changes the predominant phase of the film to this low volume phase, with very minor fraction in the m-phase (Fig. 1d). To determine the symmetry of the low-volume phase for the films with x=0.85, a three peak Gaussian fitting was performed to the pole-slicing plots from P1, P2 and P3 between 2θ of 27º and 33º, (*Supplementary Fig. S1b*). The peak position representing the low-volume phase is at 2θ=30.24±0.03º for P1, P2 and P3 whereas the out-of-plane peak is at 2θ=30.12±0.04º (see Fig. 2a-b, as well as *supplementary information,* Fig. S2, for error estimation from pole-slicing that can occur through slight misalignment and inhomogeneous strain). Such a 3:1 multiplicity in $d_{\{111\}}$ is consistent with the low-volume phase having a rhombohedral symmetry, a phase which was recently discovered on films grown epitaxially on STO substrates[39]. The $d_{\{111\}}$ and $d_{\{11\text{-}1\}}$ can be calculated to be 2.96 (±0.01) Å and 2.95 (±0.01) Å, respectively, and this corresponds to a unit cell with a=b=c≈5.11 (±0.01) Å and rhombohedral angle (α) between 89.9º and 90º. This distortion is smaller than what was observed on the 10 nm HZO (x=0.5) films grown on STO (α~89.3º)[39]. Pole figures of the {002} planes obtained around 2θ=35º and corresponding pole-slicing further confirm these lattice parameters (*Supplementary Fig. S3*).

HAADF-STEM images acquired on cross-sectional samples of films with x=0.5 clearly show that the entire film is in the monoclinic (non-polar) phase, consistent with the XRD data. EDS analysis shows a contiguous layer of amorphous a-SiO$_x$ of ~0.5 to 1 nm between Si and the HZO layer (*Supplementary Fig. S4*). This is a reformed oxide layer, a result of the backward reaction upon increasing the product concentration in the scavenging chemical reactions 1 and 2. With Zr concentration increased to x=0.7, this a-SiO$_x$ layer exists in some regions, but is absent in some other regions (Fig. 2c), a clear testimony to the better scavenging properties of Zr ions compared to Hf ions. Very interestingly, upon analyzing the $d_{\{111\}}$ lattice parameters, we find that the film just above the regions with the a-SiO$_x$ layer displays a monoclinic non-polar phase ($d_{\{11\text{-}1\}}$≈2.82 Å, $d_{\{111\}}$≈3.13 Å), while the film directly in contact with Si substrate is in the low volume phase (Fig. 2c, zoomed in Fig. 2d). HAADF-STEM multislice image (200 KV) simulations (Fig. 2e) obtained from the rhombohedral (*R3*) phase (cross-sectional sample thickness of ~20 nm), among all the other polymorphs, show the closest resemblance to our images. In particular, the alternating intensity of the cationic columns along the <112> direction is a distinct feature of the rhombohedral polar phases (both *R3* and *R3m*), which can be seen in our experimental images (Fig. 2d and 2f).

**Interfacial phase, and epitaxy:** At the interface between Si and r-phase HZO, we observe at least two monolayers of crystalline phase, which is different from both the Si and HZO structures, as shown in the bright field STEM image in Fig. 3a (contrast digitally inverted). This is most likely a crystalline, tridymite or β-cristobalite, phase of SiO$_2$, which has been studied in detail at the Si/a-SiO$_x$ interfaces, and is well-known to induce epitaxy between Si and YSZ layers[50,61–63]. In Fig. 3b, we propose a rough schematic for epitaxy, based on Fig. 3a, assuming the lattice parameters of the β-cristobalite as the crystalline c-SiO$_2$ phase (in plane~3.55Å). Epitaxy of HZO (in-plane ~3.6 Å) on c-SiO$_2$ thus provides initial compressive strain boundary conditions.

These results enable us to propose that the phase of the initial seed formed as a result of scavenging the native oxide is the low-volume r-phase. The compressive strain conditions stabilize the r-phase, quite similar to HZO films grown on STO[39]. However, regrowth of amorphous oxide layer relaxes the strain, stabilizing the non-polar m-phase[50]. Increasing the Zr content increases the efficiency of the native oxide scavenging, thus stabilizing a greater volume of the film in the r-phase.

**Polar nature of the r-phase:** Metal-insulator-semiconductor (MIS) capacitors were fabricated on these films with TiN (200 μm diameter) as the top-electrode. These were quite leaky especially in the inversion regime of operation. Cooling the devices to 10 K reduced the leakage, despite not completely avoiding it. On the films with x=0.7, very weak polarization switching (evident from I-V loop) was observed with ΔP$_s$ of ~2 μC/cm$^2$ (Fig. 3c), with devices tested at 1000 Hz. While the device optimization and rigorous electrical characterization is a subject of future work, these measurements demonstrate that the r-phase is indeed polar. About the weak ferroelectric switching, apart from the fact that these samples show a predominant fraction of non-polar m-phase, it must also be noted that the rhombohedral angle in our samples is very close to 90°. First-principles calculations[39] show that at these low distortions, the phase with *R3m* symmetry has very low Ps~1 μC/cm$^2$.

## (0.3) HfO$_2$- (0.7)ZrO$_2$ on Si (100)

**Textured {100} films, and thickness dependence:** On 5 nm thick films (Fig. 4a-b), pole figures about 2θ=31° (approximately d$_{\{111\}}$ of HZO) show 4 {111} poles at χ≈53° separated in φ by 90° (Fig. 4a). Combined with the pole figure data at 2θ=34°, which shows a {200}-pole at χ≈0° (Fig. 4b), it is evident that these films are strongly textured with {100} planes parallel to the Si (100) surface. On 10 nm thick films (Fig. 4c,d), along with all the poles observed for the 5 nm films, some extra satellites appear. Specifically, two satellites appear at χ≈58° for every intense {111} pole (at χ≈53°) separated in φ by ~7° (Fig. 4c). About the intense center {100} pole, 4 different satellites appear centered at χ≈8° (with a spread of 8-10°) separated in φ by 90° (Fig. 4d). These satellites hint of domains/grains where the {100} planes are misoriented with respect to Si (100) surface. As the thickness of the film increases to 20 nm (Fig. 4e,f), poles from these misoriented grains become the most intense and the signal from the domains with {100}//Si (100) domains almost disappears.

**Phase coexistence, domains and accordions:** To further learn about the lattice parameters and global symmetries, pole-slicing was performed across each of these pole figures. Here we discuss the results from 10 nm films on Si (100), which include all the features of both the 5 nm and the 20 nm films. Around the {111} pole (zoomed in view in Fig. 5a), when aligned perfectly either at $\chi \approx 53°$ (pixel 1 (black) in Fig. 5a), or at the most intense spot of the satellites ($\chi \approx 58°$, grey pixels 10 and 11 in Fig. 5a), pole-slicing clearly shows two distinct peaks centered at $2\theta=28.5°$ and $31.5°$ (Fig. 5b). This is a signature of the monoclinic phase, and its corresponding domain structure. In particular, the monoclinic domains that give rise to these Bragg peaks (at $\chi$ and $\varphi$ of the pixels 1, 10 and 11 in Fig. 5b), correspond to two types of domains: those having the a-b plane//Si(001) (dominating in thinner samples), and those with the a-b plane a few degrees misoriented from Si (001) (in thicker samples), as sketched n Fig. 5c. The separation in $\varphi$ of the satellites is a result of domains with positive and negative misorientation. Rigorous proof of this model through a complete mathematical analysis of the pole figure analysis is provided in the supplementary information (*Supplementary Fig. S5, 6 and related text*).

When pole slicing is performed at $\chi$ values in between the black and grey pixels (purposely "misaligned") in Fig. 5a (pixels 4,5 (red) and 3,4 (orange) pixels), we find the emergence of a third peak at $2\theta \approx 30.2°$ in addition to the two monoclinic {111} peaks (Fig. 5b). This corresponds to a low-volume minor phase that coexists with the major bulk-monoclinic phase.

Overview cross-sectional HAADF-STEM image acquired along the [010] zone (Fig. 6a) on the 10 nm thick film clearly shows an accordion-like domain morphology. Upon further zooming in (blue box in Fig. 6a), we see the domains that contribute to the zig-zag pattern correspond very well to the monoclinic symmetry (HAADF-STEM simulation in Fig. 6b). The monoclinic angle ($\beta$) varies between $81°$ and $85°$ across various domains. This variation is indeed reflected as the spread of the satellite spots about $\chi=8°$ in the {001} pole figure (Fig. 4d). Intersection of two domains with positive and negative misorientation (9 and -9° in Fig. 6b) of the ab plane with Si (001), results in the accordion morphology, consistent with the satellite spots of the {111} poles in XRD analysis. The monoclinic domains corresponding to ab plane//Si (001) are shown in the supplementary information (*Supplementary Fig. S7*).

The region enclosed in red in Fig. 6a, when zoomed and further analysed is consistent with our HAADF-STEM simulations from an orthorhombic crystal (Fig. 6c), and not the other low-volume polymorphs (simulations of various polymorphs are compared in *Supplementary Fig.S8*). Interplanar (d) spacings for various crystallographic planes obtained from the HAADF-STEM image Fourier transform (Fig. 6c), match very well with o-phase bulk lattice parameters[64]. In particular, $d_{(in-plane)}$=2.51±0.02 Å, and $d_{(out-of-plane)}$=2.57±0.02 Å, correspond to $d_{(002)}$ and $d_{(200)}$ for the o-phase, suggesting that the polar axis (*c*-axis in *Pca2₁* symmetry) is in-plane. Thus STEM analysis conclusively shows that, in addition to the various kinds of monoclinic domains (Fig. 5c), a low-volume o-phase, the commonly occurring low-energy ferroelectric phase in polycrystalline thin films, is also present in these films.

**Interface and epitaxy:** From EDS chemical maps (Fig. 7a) and Wiener filtered HAADF-STEM images (Fig. 7b) at the interface between HZO and Si (100), it is quite clear that a contiguous layer of <1 nm regrown a-$SiO_x$ exists. Furthermore, there also seems to be an interface c-$SiO_2$ (β-cristobalite, most likely) structure between Si(100) and the a-$SiO_x$, which enables epitaxy on Si(001)[50,61–63]. The in-plane lattice parameter of β-cristobalite (5.03 Å) imposes a compressive strain on any polymorph of HZO. However, the regrowth of a-$SiO_x$ relaxes this strain stabilizing most of the film in a bulk m-phase. Notably, the grains stabilizing in the o-phase also are interfaced with regrown a-SiOx. This is unlike growth of HZO (x=0.7) on Si (111), where there is a clear correlation between the existence of a low-volume phase, and the absence of regrown a-$SiO_x$ on Si. Thus, it appears that the stabilization of o-phase is a result of the inhomogeneous strain fields originating at the intersection of various kinds of nanoscopic monoclinic domains that form the accordion, and not because of strain transfer from the substrate.

**Polar r-phase vs. polar o-phase**

Among all the polar polymorphs of hafnia, first principles calculations suggest that the o-phase (*Pca2₁*) has the least energy (64 meV f.u.$^{-1}$ with respect to the ground state)[39]. The rhombohedral polymorphs (*R3m*: 158 meV f.u.$^{-1}$ and *R3*: 195 meV f.u.$^{-1}$ ), though more energetic, seem more favorable to obtain experimentally, under a combination of compressive strain and quite notably a (111) film orientation. Both the films grown on STO substrates by Wei et al.,[39] and the films grown on Si(111) in the current work satisfy these conditions, and thus exhibit a polar r-phase (and not the low-energy o-phase). These observations, however, are quite contradictory to the theoretical predictions of Liu and Hanrahan[22], possibly owing to the absence of the r-phases in their calculations. It must also be noted that films grown in (001) orientation in this work, show a preference for the o-phase, and not the r-phase. Such an orientation dependence of the obtained polar phase is very unique, and deserves further investigation.

**Conclusions:**

(1-x)HfO$_2$-(x)ZrO$_2$ (HZO) with various compositions in the range 0.5< x< 0.85 were grown epitaxially on Si(111) and Si(100), without using buffer layers, using PLD. *In situ* scavenging of the native oxide using decomposition reactions plays a crucial role in achieving epitaxy directly on Si. On both these substrates, an interfacial phase of c-SiO$_2$ (most likely β-cristobalite) has been found, and it offers initial compressive strain conditions for the growth of HZO. The regrowth of an amorphous a-SiO$_x$ interlayer is a relaxation mechanism to release strain. On Si (111), the film on the top of the regions of regrown amorphous oxide relaxes to a bulk non-polar monoclinic phase, and the film directly in connection with the c-SiO$_2$ is in the polar r-phase. The volume fraction of the r-phase increases and the regrown a-SiO$_x$ decreases with increasing Zr content, owing to better reactivity of Zr to participate in the a-SiO$_x$ scavenging reaction (compared to Hf). Ferroelectric measurements show leaky and incipient, but clear, P-E hysteresis loops, an evidence to the polar nature of the r-phase. On Si (100), the observed polar o-phase seems to be stabilized by inhomogeneous strains arising out of nanodomain coexistence of the surrounding m-phase. Finally, in addition to strain and surface energy, there also appears to be an orientational dependence to the stabilization of polar phases (at least on Si), i.e. the r-phase is favored in (111) orientation, while the o-phase is preferred in (001).

**Acknowledgements:** PN acknowledges the funding received from European Union's Horizon 2020 research and innovation programme under Marie Sklodowska-Curie grant agreement No: 794954 (Project name: FERHAZ). JA and BN acknowledge the funding from NWO's TOP-PUNT grant 718.016002. YW and BN acknowledge the China Scholarship Council. LY and BD acknowledge a public grant overseen by the French National Research Agency (ANR) as a part of the `Investissements d'Avenir' program (grant no. ANR-10-EQPX-37, EquipEx MATMECA) and through ANR-17-CE24-0032/EXPAND.

**Figure Captions**

**Figure 1. Texture and phase analysis on 10 nm (1-x)HfO$_2$: xZrO$_2$ films grown on Si (111)**

(a) Representative pole figure obtained about 2θ=30° on film with x=0.7, consistent with <111> out of plane texture. The three non-out-of-plane {111} poles, P1-P3, arising out of the film are centered at χ≈71° and separated in φ by 120°. Rotated 180° from the film pattern, we see weak spots arising from the tail of substrate non-out-of-plane {111} poles at 2θ=28.44°. Pole figure symmetry looks similar for x=0.5 and x=0.85 (Supplementary Fig. 1). (b-d) Pole-slicing, or θ-2θ scans around P1, P2 and P3 for x=0.5 (b), 0.7 (c) and 0.85 (d). While x=0.5 (b) shows just the monoclinic {111} peaks, a low-volume phase peak starts evolving from x=0.7 (c) at 2θ≈30,2° and intensifies at x=0.85 (d).

**Figure 2. Determination of symmetry for the low-volume phase on Si (111)**

(a) Gaussian fits for the non-out-of-plane {111} poles P1-P3 and the plane normal (out-of-plane) corresponding exclusively to the low-volume phase for films with x=0.85. The raw data (Fig. 1d), and 3 peak fitting procedure (two monoclinic, and one low-volume phase) is shown in Supplementary Fig. 1b. (b) Peak positions of the P1-P3 compared to the plane normal obtained from (a). This clearly reveals a 3:1 multiplicity, or an r-phase. Error estimation is shown in Supplementary Fig.2. r-phase is further corroborated from pole-slicing of the {001} poles shown in Supplementary Fig. 3. (c) Cross-sectional HAADF-STEM image of TiN-HZO (x=0.7)-Si MIS capacitor (obtained along the <1-10> zone of the substrate), showing various {111} d-spacings. $d_{\{11-1\}}$=2.81±0.02 Å, and $d_{\{111\}}$=3.11±0.02 Å are quite clearly bulk m-phase parameters. The low-volume phase becomes a suspect when $d_{\{111\}}$ values are measured between 2.90 and 3.00 Å. (d) Zoomed-in look at the Si-HZO interface, boxed in yellow in (c). The interface with Si on the left contains re-grown a-$SiO_x$ (<1 nm). We clearly see an m-phase right above it. The region on the right has no a-$SiO_x$, but instead a direct connection between HZO and the crystalline substrate. That is a low-volume phase. (e) R3 phase HAADF-STEM multislice image simulation at lamella thickness of ~20 nm (defocus ~0). This quite corresponds to our image. In particular the contrast fluctuations along the red line (<112> direction) in (d) are shown in (f), do not appear in o or t-phase, and are quite unique to the r-phases (as can be seen visually in (e)).

**Figure 3. Interface structure and epitaxy on Si (111)**

(a) Bright field (BF) cross-sectional STEM image with inverted contrast of the interface. BF mode provides better contrast for lighter elements, and thus this mode of imaging is being reported. Clearly, we see 2-3 monolayers of interface with a different structure than both Si substrate and the HZO. This is a c-$SiO_2$ interface, with β-cristobalite or tridymite being the most likely phases. (b) Schematic of epitaxy on Si, with β-cristobalite at the interface. It provides the compressive strain necessary to stabilize the r-phase on HZO. (c) I-V, and corresponding P-V loops obtained from MIS capacitor structure of TiN-HZO (x=0.7)-Si at 1000 Hz, and 10 K. Although quite leaky on the inversion side (negative voltage), we nevertheless see switching peaks in the I-V curves. $P_s$ is quite low, since the majority of the film is in a non-polar monoclinic phase. But this is a proof-of-concept that the r-phase is indeed polar.

**Figure 4. Texture measurements on HZO (x=0.7) on Si(100) at different thicknesses**

Pole figures obtained about 2θ=31º (corresponding to $d_{\{111\}}$) and about 2θ=34º (corresponding to d{100}) on 5 nm (a,b), 10 nm (c,d) and 20 nm (e,f) HZO films grown on Si (100). (a,b) In 5 nm thick samples 4 {111} poles separated in φ by 90º are observed at χ≈53º, and {001} pole at χ≈0º clearly showing that the films are textured with <001> oriented out of plane. (c,d) In 10 nm thick samples, apart from the poles observed for 5 nm, there are some extra satellite {111} poles at χ≈58º forming a triangular pattern, and 4 {001} satellites centered around χ≈8º (spread=8-10º) separated in φ by 90º. All these patterns can be mathematically explained through the coexistence of various monoclinic domains (each contributing to a different Bragg reflection, see Supplementary information). (e,f) What were just satellite spots in 10 nm samples, become the main Bragg spots at 20 nm thickness.

**Figure 5. Phase coexistence and domains**

(a) From figure 4c (10 nm film thickness), one of the {111} poles and its corresponding satellites are zoomed in. The black pixel corresponds to $\chi = 53°$, where the {111} main pole is the most intense. The grey pixels correspond to $\chi = 58°$ where the satellites are the most intense. A progression from black to grey traverses through orange and red pixels, which correspond to purposeful misalignment ("well aligned" conditions are obtained by maximizing the intensity, which would ignore information from orange and red pixels). (b) With the color coding scheme as described in (a), pole slicing was then performed at various $\chi$ values from black pixel to the grey pixel (pixel positions numbered 1 to 11). At the black (1) or grey (10, 11) pixels, peaks corresponding to monoclinic {111} d-spacings are observed. In the orange (2, 3) and red (4, 6) pixels however, a {111} peak corresponding to a low-volume phase appears at $2\theta \approx 30.2°$. (c) The various domains that correspond to the pole-figures presented in Figure 4. The left one shows a monoclinic domain with a-b plane of HZO// Si(100). This is very well oriented growth observed at low film thicknesses (5 nm). The center panel shows a monoclinic domain with a-b plane misaligned from the Si (100) surface. Negative and positive misorientations can give rise to an accordion-like zig-zag pattern. The right most panel shows the domain from the low-volume phase (which can be shown to be orthorhombic from TEM results to follow)

**Figure 6. Monoclinic domains, orthorhombic domains, and accordions of HZO on Si (100)**

(a) Overview cross-sectional HAADF-STEM image along the [010] zone axis, of 10 nm thick HZO (x=0.7), showing a clear accordion-like morphology. (b) Zoomed-in view of the region surrounded by blue box in (a), showing intersection of two domains. D-spacings estimated from the Fast Fourier transform (FFT) obtained from the domain on the left confirm a monoclinic phase. Furthermore, the monoclinic angle ($\beta$) can be directly read from the FFT to be ~81° in this image. The ab plane is misoriented with respect to the substrate (normal) by 9°. HAADF-STEM multislice image simulations of monoclinic phase, clearly match our domain. If the domain on the left is positively misoriented with the substrate, the domain on the right is negatively misoriented, and their intersection gives rise to an accordion pattern. (c) Zoomed-in view of the region surrounded by the red box in (a). The d-spacing obtained from FFT clearly shows that this is in the o-phase ($d_{(002)}=2.51\pm0.02$ Å, and $d_{(200)}=2.57\pm0.02$ Å). HAADF-STEM image simulations match the real image (with zig-zag arrangement of atomic columns along [001]). Simulations from different phases are compared in supplementary Fig. 5. The polar axis however, is the c-axis which is in-plane.

**Figure 7. Interface and epitaxy on Si (100)**

(a,b) EDS chemical maps, and Wiener-filtered HAADF-STEM image clearly showing a re-grown a-SiO$_x$ layer of ~ 1 nm between Si (100) and HZO (x=0.7). (c) Zoomed-in view of Wiener filtered image in (b). There is also an interfacial crystalline structure (boxed in blue) at the interface of Si and a-SiO$_x$, most likely belonging to c-SiO$_2$ in β-cristobalite phase. This enables orientational relationship with HZO layer.

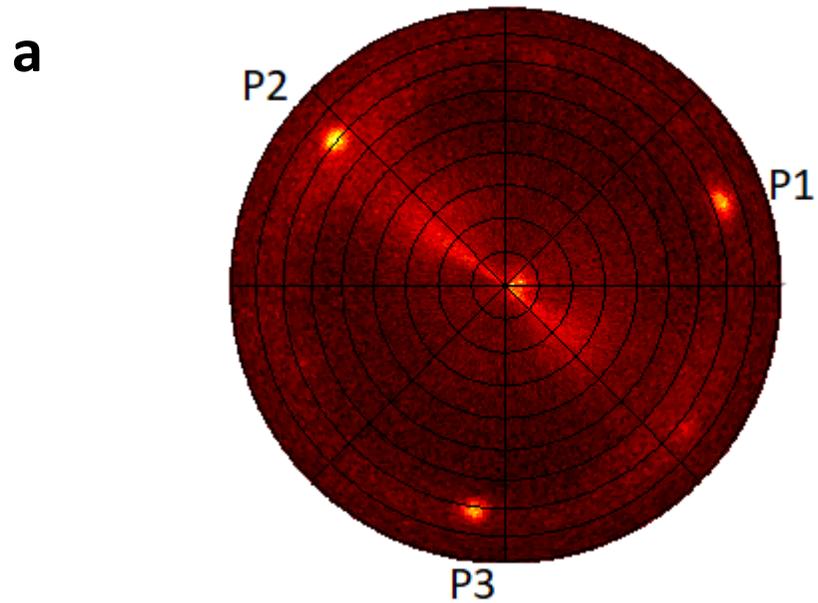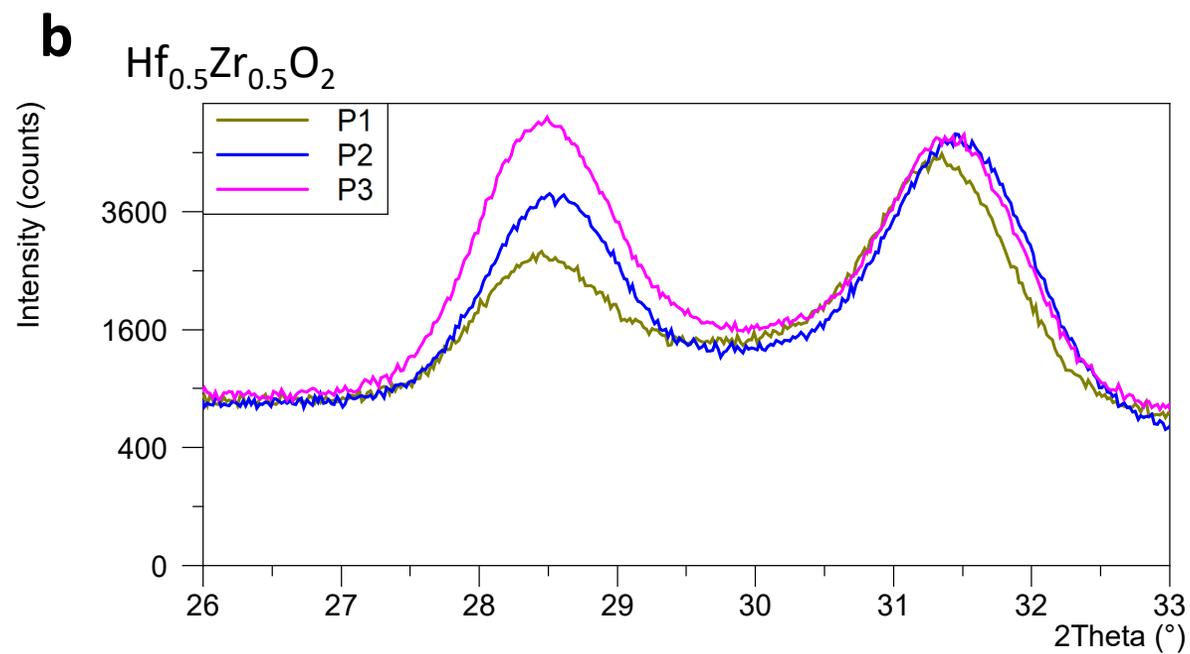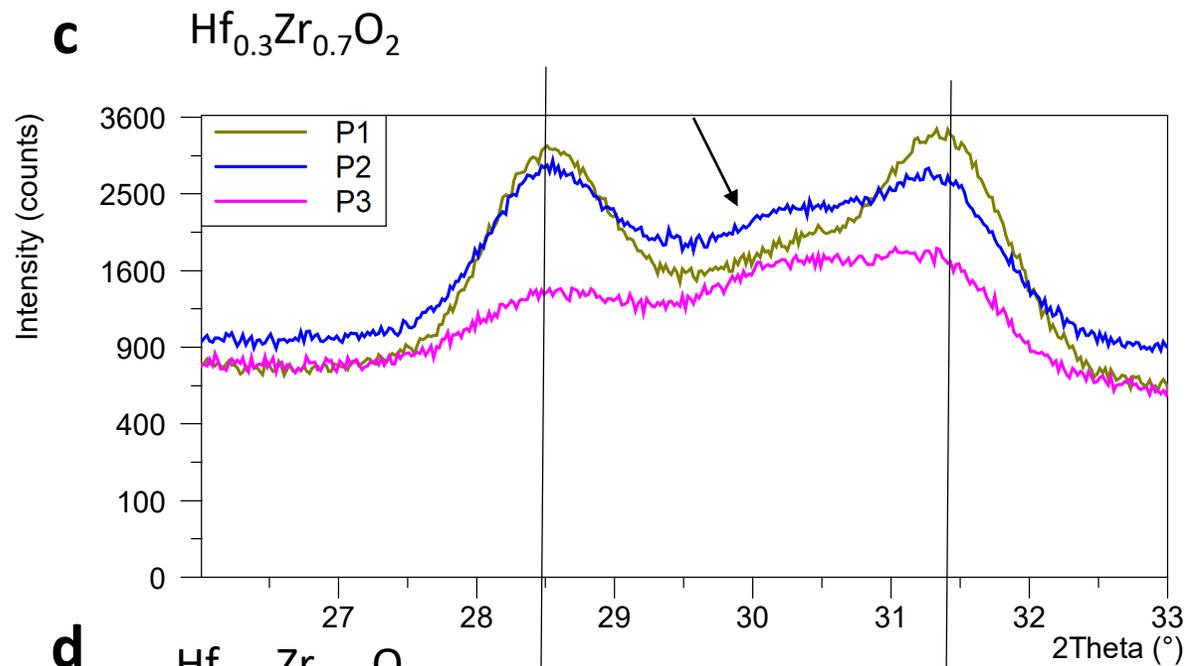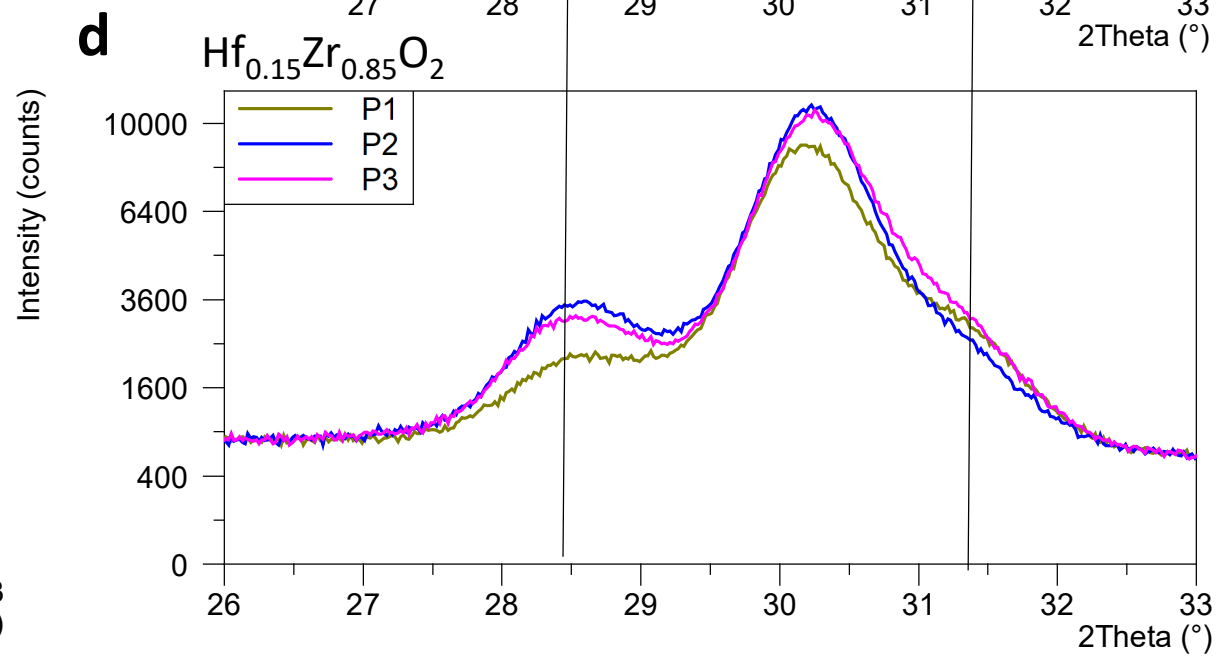

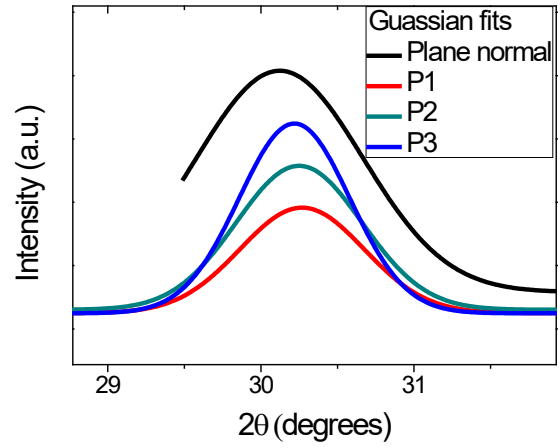
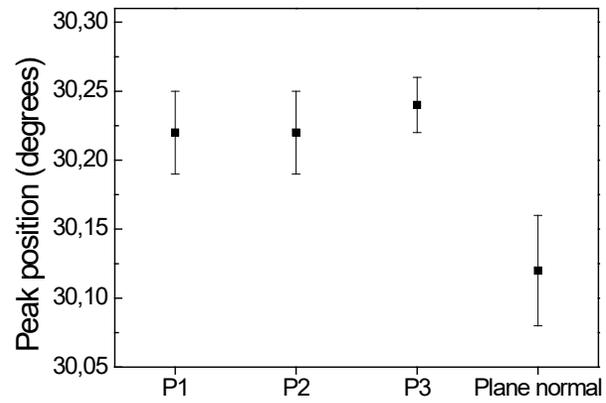
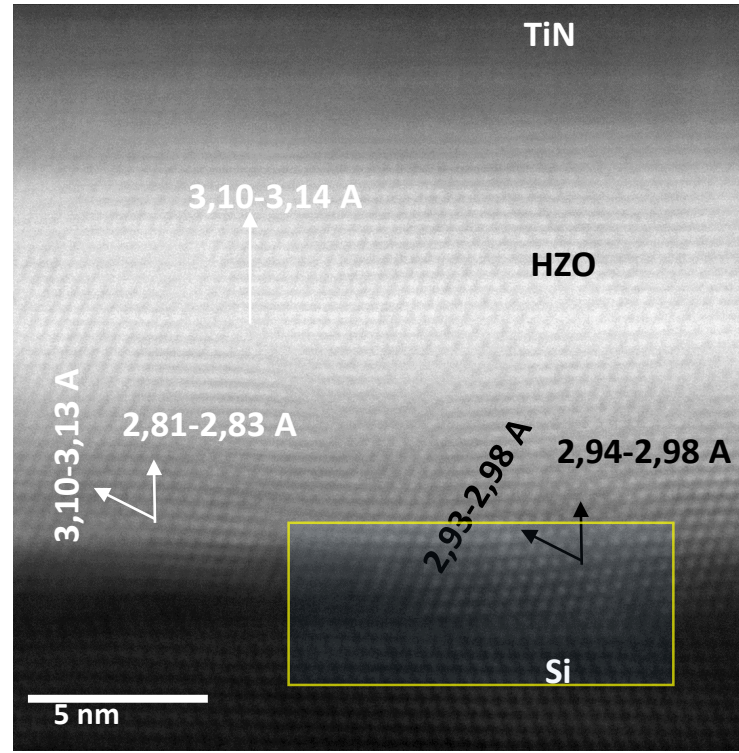
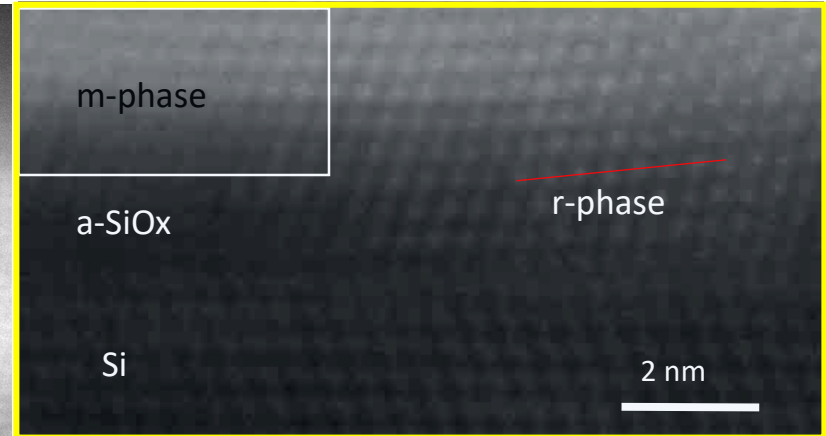
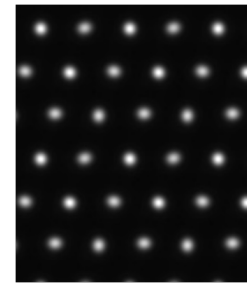
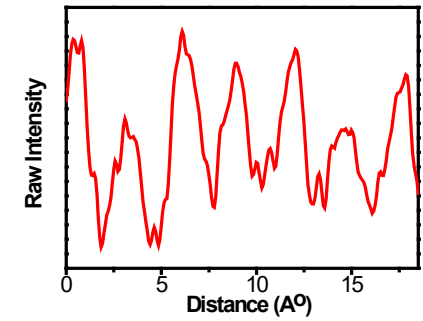

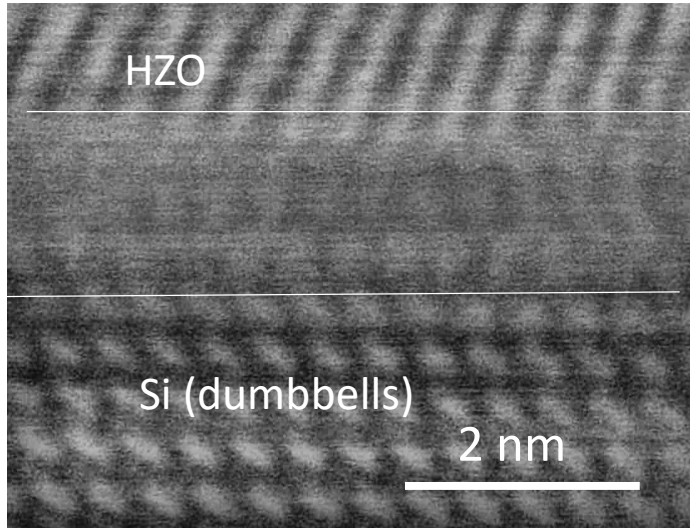 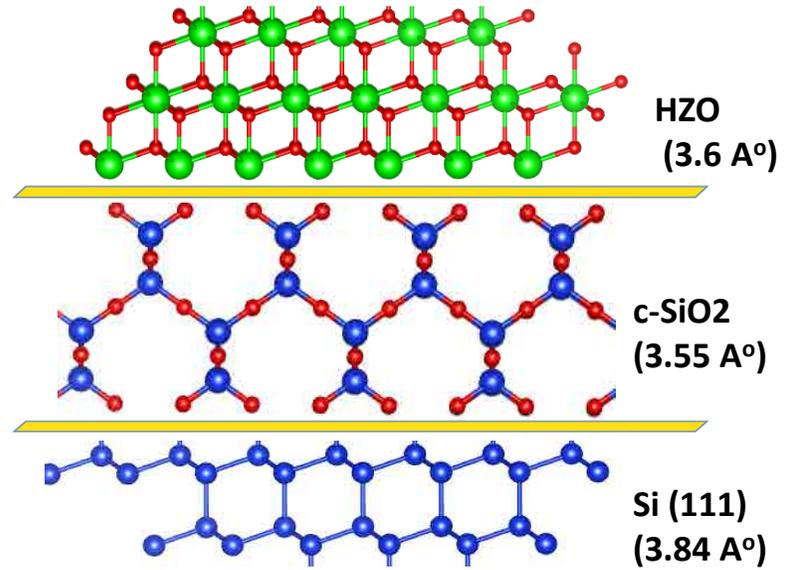 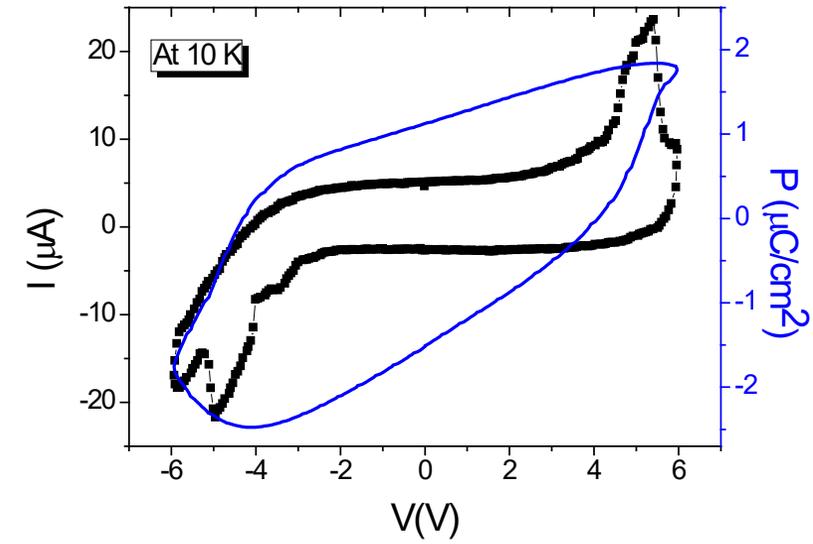

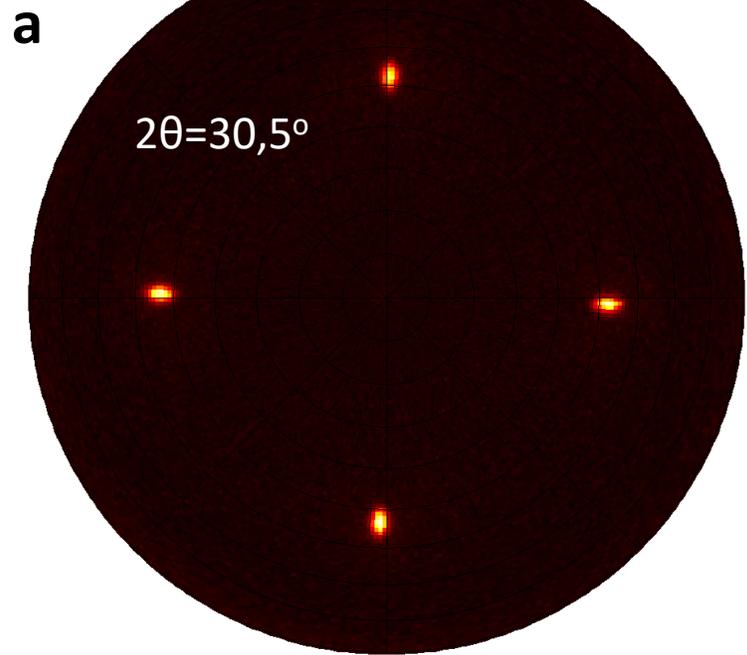
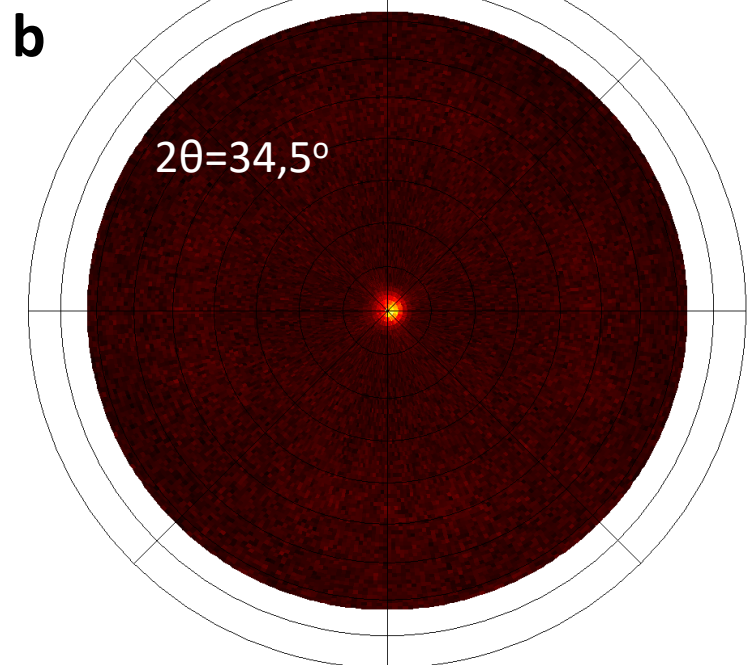
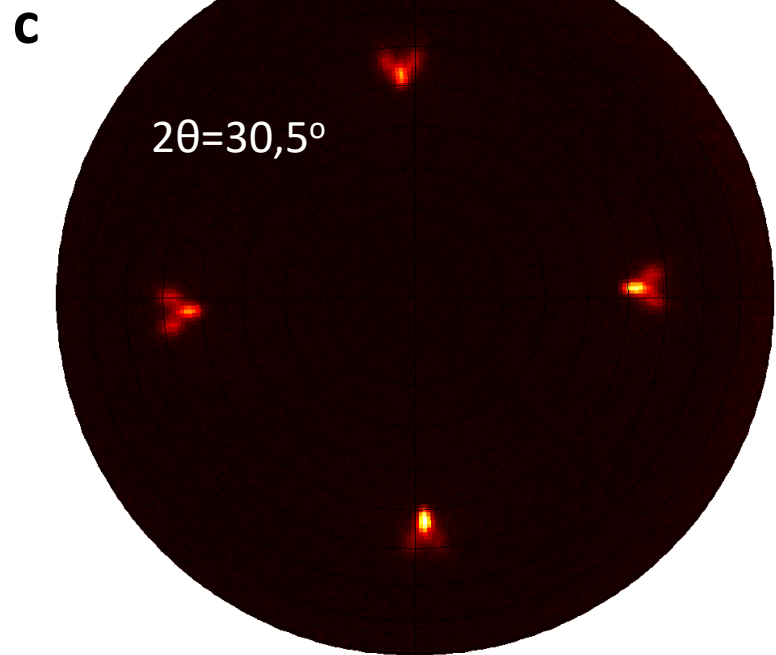
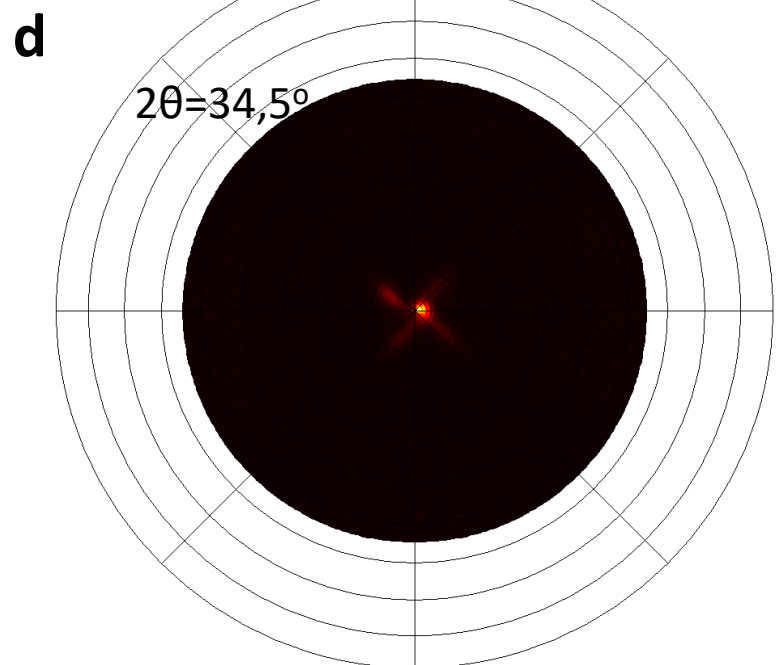
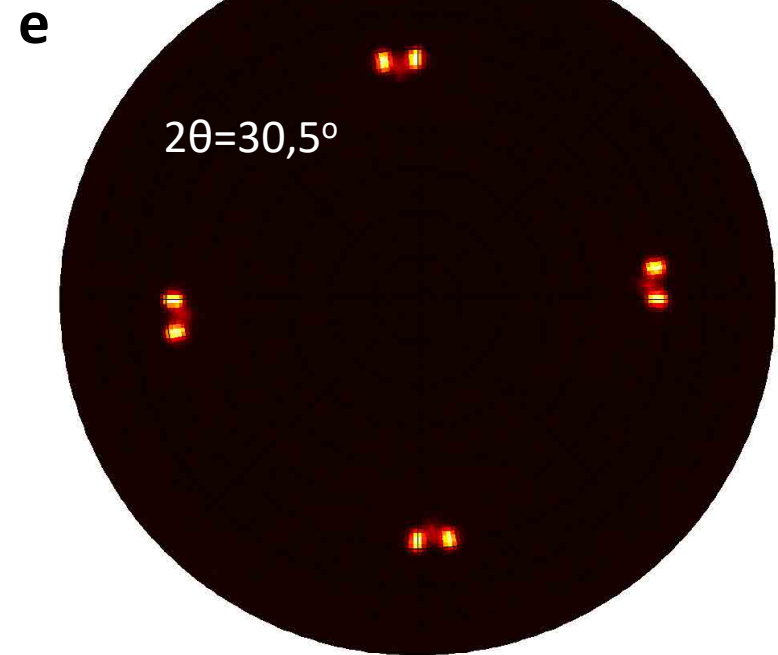
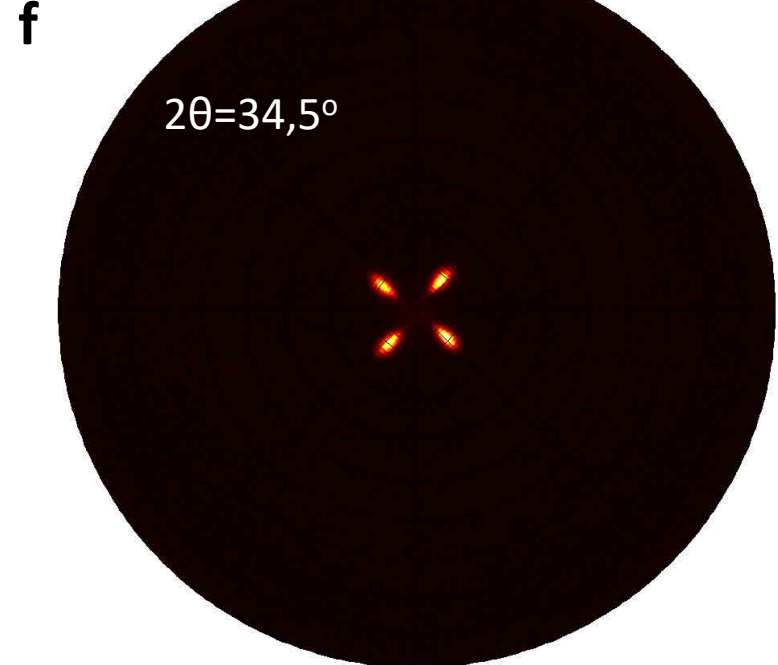

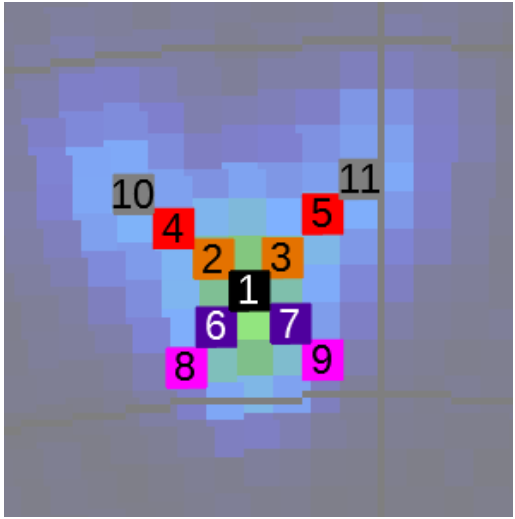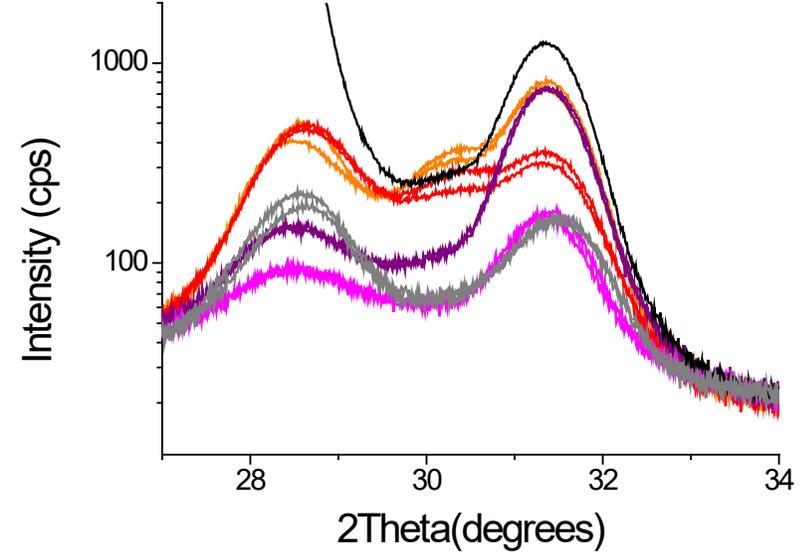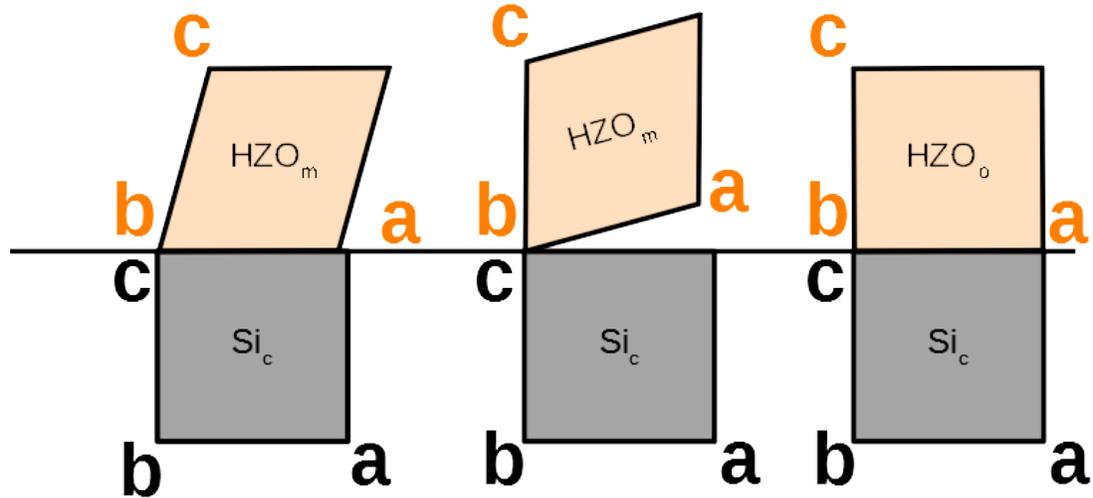

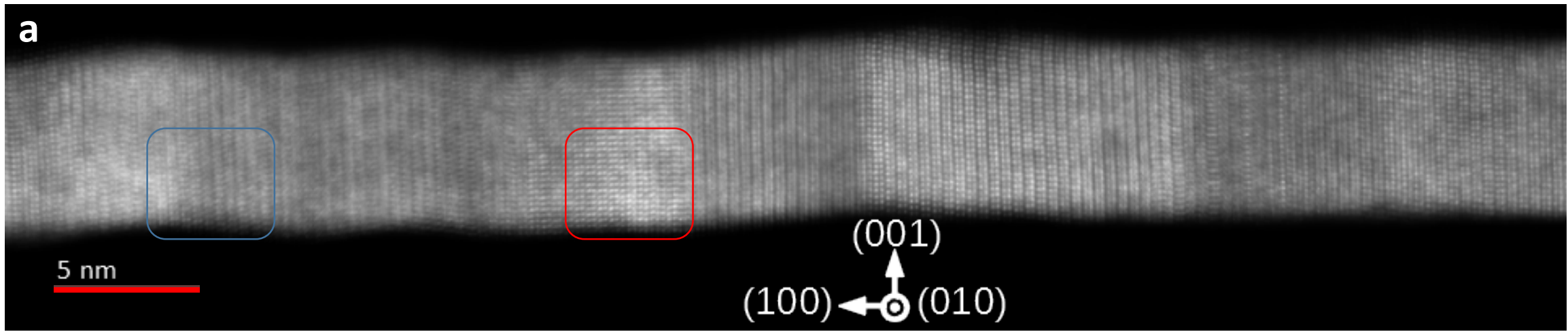
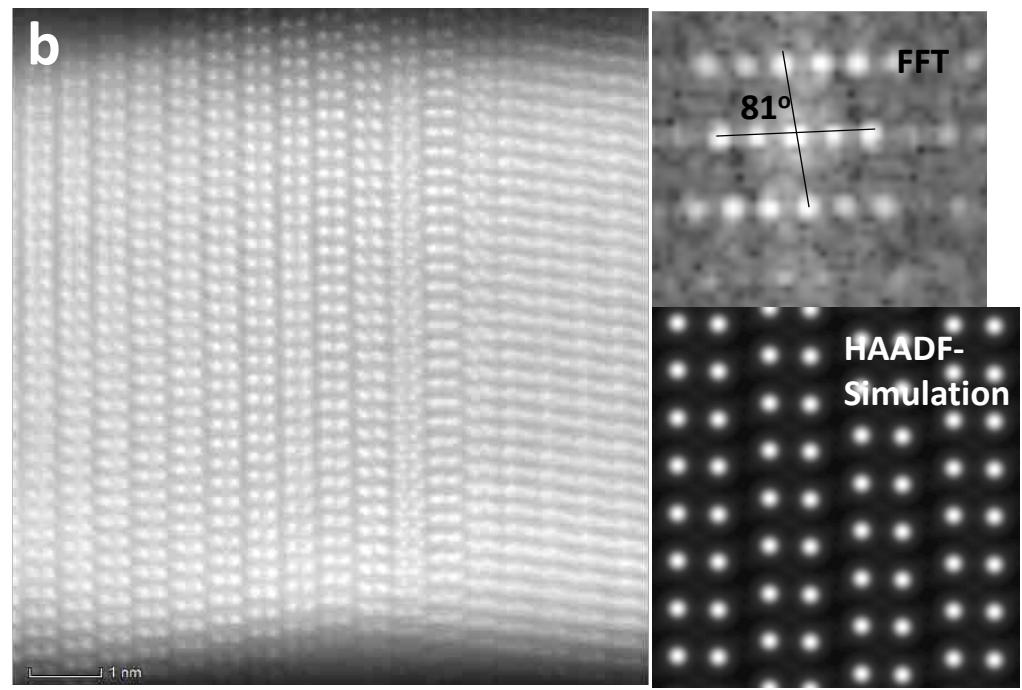
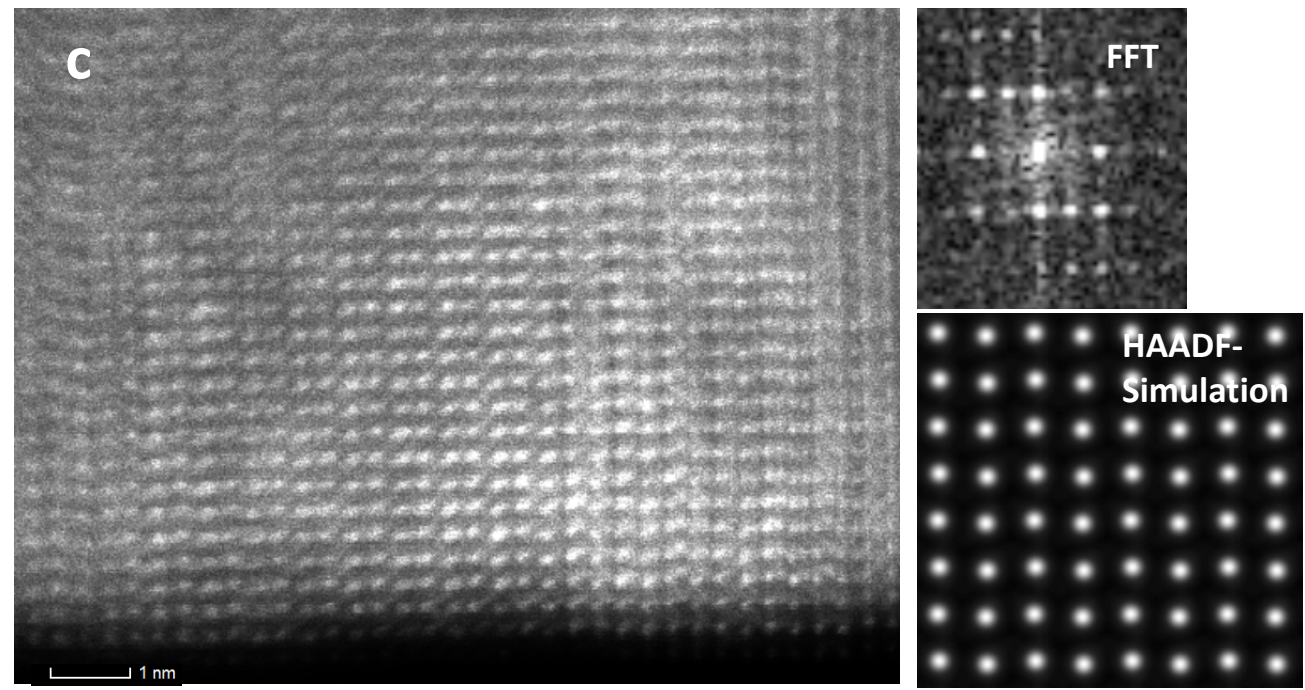

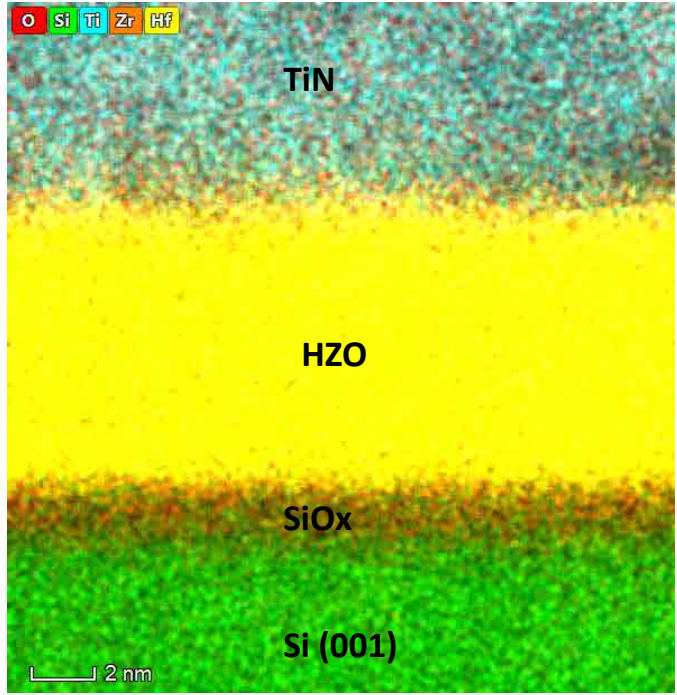 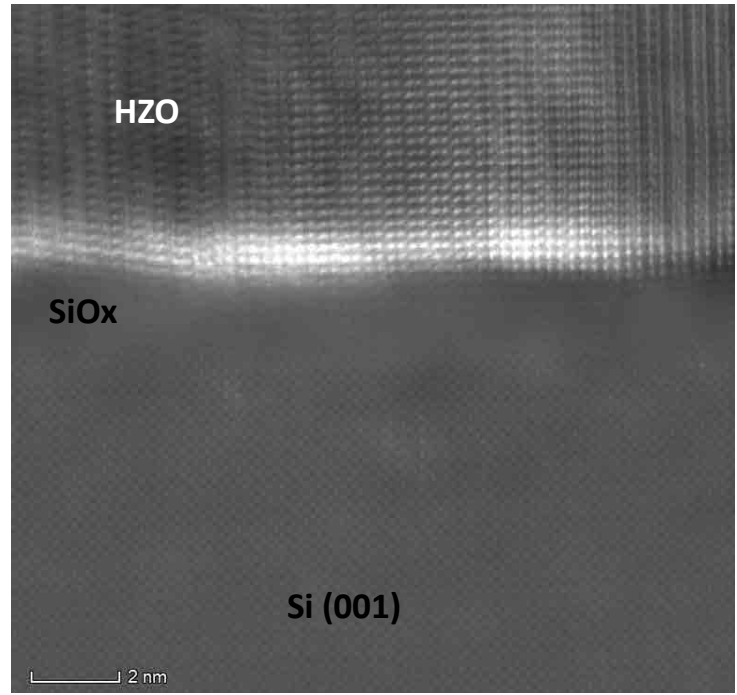 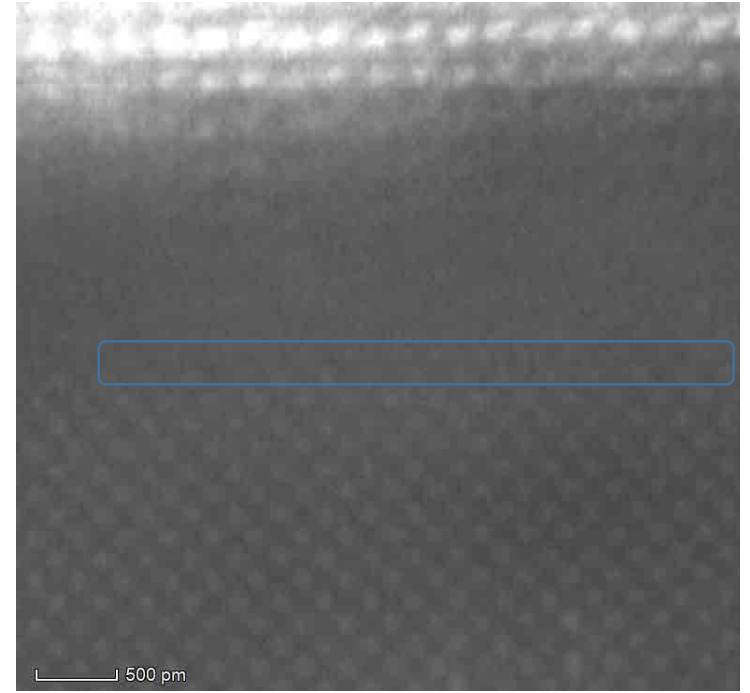